\documentstyle{mn}
\newread\epsffilein    % file to \read
\newif\ifepsffileok    % continue looking for the bounding box?
\newif\ifepsfbbfound   % success?
\newif\ifepsfverbose   % report what you're making?
\newdimen\epsfxsize    % horizontal size after scaling
\newdimen\epsfysize    % vertical size after scaling
\newdimen\epsftsize    % horizontal size before scaling
\newdimen\epsfrsize    % vertical size before scaling
\newdimen\epsftmp      % register for arithmetic manipulation
\newdimen\pspoints     % conversion factor
\def\epsfbox#1{\global\def\epsfllx{72}\global\def\epsflly{72}%
   \global\def\epsfurx{540}\global\def\epsfury{720}%
   \def\lbracket{[}\def\testit{#1}\ifx\testit\lbracket
   \let\next=\epsfgetlitbb\else\let\next=\epsfnormal\fi\next{#1}}%
\def\epsfgetlitbb#1#2 #3 #4 #5]#6{\epsfgrab #2 #3 #4 #5 .\\%
   \epsfsetgraph{#6}}%
\def\epsfnormal#1{\epsfgetbb{#1}\epsfsetgraph{#1}}%
\def\epsfgetbb#1{%
\openin\epsffilein=#1
\ifeof\epsffilein\errmessage{I couldn't open #1, will ignore it}\else
   {\epsffileoktrue \chardef\other=12
    \def\do##1{\catcode`##1=\other}\dospecials \catcode`\ =10
    \loop
       \read\epsffilein to \epsffileline
       \ifeof\epsffilein\epsffileokfalse\else
          \expandafter\epsfaux\epsffileline:. \\%
       \fi
   \ifepsffileok\repeat
   \ifepsfbbfound\else
    \ifepsfverbose\message{No bounding box comment in #1; using defaults}\fi\fi
   }\closein\epsffilein\fi}%
\def\epsfsetgraph#1{%
   \epsfrsize=\epsfury\pspoints
   \advance\epsfrsize by-\epsflly\pspoints
   \epsftsize=\epsfurx\pspoints
   \advance\epsftsize by-\epsfllx\pspoints
   \epsfxsize\epsfsize\epsftsize\epsfrsize
   \ifnum\epsfxsize=0 \ifnum\epsfysize=0
      \epsfxsize=\epsftsize \epsfysize=\epsfrsize
     \else\epsftmp=\epsftsize \divide\epsftmp\epsfrsize
       \epsfxsize=\epsfysize \multiply\epsfxsize\epsftmp
       \multiply\epsftmp\epsfrsize \advance\epsftsize-\epsftmp
       \epsftmp=\epsfysize
       \loop \advance\epsftsize\epsftsize \divide\epsftmp 2
       \ifnum\epsftmp>0
          \ifnum\epsftsize<\epsfrsize\else
             \advance\epsftsize-\epsfrsize \advance\epsfxsize\epsftmp \fi
       \repeat
     \fi
   \else\epsftmp=\epsfrsize \divide\epsftmp\epsftsize
     \epsfysize=\epsfxsize \multiply\epsfysize\epsftmp   
     \multiply\epsftmp\epsftsize \advance\epsfrsize-\epsftmp
     \epsftmp=\epsfxsize
     \loop \advance\epsfrsize\epsfrsize \divide\epsftmp 2
     \ifnum\epsftmp>0
        \ifnum\epsfrsize<\epsftsize\else
           \advance\epsfrsize-\epsftsize \advance\epsfysize\epsftmp \fi
     \repeat     
   \fi
   \ifepsfverbose\message{#1: width=\the\epsfxsize, height=\the\epsfysize}\fi
   \epsftmp=10\epsfxsize \divide\epsftmp\pspoints
   \newcount\figskipcount
      \message{#1 \the\epsfysize  }
   \vbox to\epsfysize{\vfil\hbox to\epsfxsize{%
      \includegraphics{#1}%
      \hfil}}%
\epsfxsize=0pt\epsfysize=0pt}%

{\catcode`\%=12 \global\let\epsfpercent=%\global\def\epsfbblit{%BoundingBox}}%
\long\def\epsfaux#1#2:#3\\{\ifx#1\epsfpercent
   \def\testit{#2}\ifx\testit\epsfbblit
      \epsfgrab #3 . . . \\%
      \epsffileokfalse
      \global\epsfbbfoundtrue
   \fi\else\ifx#1\par\else\epsffileokfalse\fi\fi}%
\def\epsfgrab #1 #2 #3 #4 #5\\{%
   \global\def\epsfllx{#1}\ifx\epsfllx\empty
      \epsfgrab #2 #3 #4 #5 .\\\else
   \global\def\epsflly{#2}%
   \global\def\epsfurx{#3}\global\def\epsfury{#4}\fi}%
\def\epsfsize#1#2{\epsfxsize}
\def\figinsert#1#2{\epsfbox{#1} \message{#2} }    %insert figures
\title[Faint CCD sequences -- II.]
      {Faint $UBVRI$ CCD sequences for wide-field surveys -- II.  $UBVR$
      sequences at $\delta=-30\degr$}
\author[S.M. Croom et al.]
       {Scott M.~Croom$^1$\thanks{Present address: Astrophysics Group,
      Imperial College of Science, Technology and Medicine, Blackett
      Laboratory, Prince Consort Road, London, SW7 2BZ, UK. E-mail:
      s.croom@ic.ac.uk},
      A.~Ratcliffe$^1$, Q.A.~Parker$^2$, T.~Shanks$^1$,
      B.J. Boyle$^3$\vspace{0.1cm}\\ {\LARGE and R.J. Smith$^4$}\\
        $^1$Physics Department, University of Durham, South Road, Durham, DH1 3LE,
England.\\
$^2$Anglo-Australian Observatory, Coonabarabran, NSW 2357, Australia.\\
$^3$Anglo-Australian Observatory, PO Box 296, Epping, NSW 2121, Australia.\\
$^4$Institute of Astronomy, Madingley Road, Cambridge, CB3 0HA, England.}
\begin{document}
 
\maketitle
\begin{abstract}
We present results from a continuing campaign to secure deep
multi-colour CCD sequences for photometric calibration in UK Schmidt
fields with galactic latitudes $|b|>50\degr$.  In this paper we
present $UBVR$ photometry in 12 fields and $BR$ photometry in a further 14
fields observed within UK Schmidt survey fields centred at
$\delta=-30\degr$.  Photometric errors are at the $0.05$ level at 20.2,
21, 20.5 and 20 mag for $UBVR$ sequences respectively.
Our observations were carried out with the 0.9m
Telescope at the Cerro-Tololo Inter-American Observatory.  These data
are not intended for use as highly accurate individual photometric
standards, but rather for use as sequences, using a large number of
stars to calibrate wide-area data such as photographic plates.  The
data are available electronically at {\tt
http://icstar5.ph.ic.ac.uk/$\sim$scroom/phot/photom.html}.
\end{abstract}
 

\section{introduction}

Many large astronomical survey projects are currently being carried
out, or planned over the next few years.  Wide-field instruments such
as the 2-degree Field (2dF) multi-fibre spectrographic system on the
Anglo-Australian Telescope (AAT) are now allowing astronomers to
observe many hundreds of objects at the same time.  Thus, surveys of many
thousands of objects are now possible, covering very large areas on
the sky.  Indeed, major
programs are already underway to obtain redshifts for $\sim250,000$
galaxies and $\sim30,000$ QSOs.  The candidate selection procedures for
these and many other projects are based on machine measurements of
photographic plates (for example, using APM or SuperCOSMOS), as this
is currently the only method that can deal with sufficiently large
areas.  

These large surveys are either solely magnitude limited (as in the
case of the 2dF Galaxy Survey, Maddox 1998), or also include colour
selection (for
example, in the case of the 2dF QSO Redshift Survey, Smith et
al. 1997).  It is
imperative that both
magnitudes and colours are well known in order to have an effective
selection of target objects.  This is particularly the case when the
survey area covers 
a large number of photographic plates, in order that the survey be
consistent over its entire area.

In Boyle, Shanks \& Croom (1995) (Paper I) we began an observational
program to obtain
deep ($B<21$) CCD sequences in a large number of high galactic
latitude fields ($|b|>50\degr$).  Paper I and also Smith \& Boyle (1998)
(Paper III) are concerned with fields at
$\delta=0\degr$.  In this paper we extend this program to a
strip of UK Schmidt Telescope (UKST) survey fields at
$\delta=-30\degr$, near the South Galactic Pole.  These extend from
$21^h51$ to $03^h04$ in Right Ascension,  covering UKST survey fields
466-471 and 409-417.  Here we present the first of the observations in
our Southern fields, which consists of 12 fields with $UBVR$
photometry and a further 14 fields with $BR$ photometry.  We have
endeavoured to make this data available to the astronomical community
as soon as
possible after observation.  Therefore, computer readable data files are
available via the WWW at {\tt
http://icstar5.ph.ic.ac.uk/$\sim$scroom/phot/photom.html}.

\section{Observations}

\begin{table*}
\baselineskip=20pt
\begin{center}
\caption{Data on all CCD sequence fields.  Exposure times in
each band are listed, along with the aperture corrections used to
correct the photometric zero-points from 10 pixel to 25 pixel
apertures ($\Delta m_{\rm ap}$).  $B_{\rm lim}$ is a limiting magnitude
in the $B$ band for each field, defined as the magnitude at which
photometric errors are 0.05mag.  Fields marked $*$ were calibrated
using shorter exposures on different nights.}
\vspace{0.5cm}
\begin{tabular}{cccccccccccccc}
\hline 
Field & R.A.& Dec. & 
& & & & & & & & &\\ 
Name & (J2000) & (J2000) & Date Obs. 
& $u$ & $\Delta u_{\rm ap}$ & $b$ & $\Delta b_{\rm ap}$ & $v$ &
$\Delta v_{\rm ap}$ & $r$ & $\Delta r_{\rm ap}$ & $B_{\rm lim}$\\ 
\hline 
f409-1 &  00 01 12.78 & --27 33 48.6  & 25/08/95  &  -- & -- & 1$\times$1200 & 0.039 & -- & -- & 1$\times$600 & 0.038 &21.1 \\ 
f409-2 &  00 02 56.49 & --29 22 16.6  & 15/09/96  &  3$\times$1200 & 0.042 & 2$\times$600 & 0.051 & 2$\times$300 & 0.036 & 2$\times$300 & 0.041 &20.8 \\ 
f410-2 &  00 21 52.31 & --29 48 21.6  & 30/09/97  &  3$\times$1200 & 0.028 & 3$\times$400 & 0.027 & 2$\times$300 & 0.029 & 2$\times$300 & 0.031 &20.9 \\ 
f410-1 &  00 36 43.89 & --29 31 34.5  & 26/08/95  &  -- & -- & 1$\times$1200 & 0.068 & -- & -- & 1$\times$600 & 0.059 &21.0 \\ 
f411-2 &  00 49 47.87 & --29 09 40.0  & 28/09/97  &  3$\times$1200 & 0.022 & 3$\times$400 & 0.024 & 2$\times$300 & 0.021 & 2$\times$300 & 0.027 &20.9 \\ 
f411-1 &  00 51 19.84 & --28 30 20.4  & 26/08/95  &  -- & -- & 1$\times$1200 & 0.035 & -- & -- & 1$\times$600 & 0.035 &21.2 \\ 
f412-1 &  01 16 48.89 & --28 35 06.6  & 26/08/95  &  -- & -- & 1$\times$600 & 0.038 & -- & -- & 1$\times$300 & 0.049 &20.8 \\ 
f413-1 &  01 34 23.38 & --31 38 30.4  & 26/08/95  &  -- & -- & 1$\times$600 & 0.053 & -- & -- & 1$\times$300 & 0.056 &20.8 \\ 
f413-2 &  01 34 41.60 & --29 36 39.9  & 28/09/97  &  3$\times$1200 & 0.023 & 3$\times$400 & 0.026 & 2$\times$300 & 0.027 & 2$\times$300 & 0.028 &20.9 \\ 
f414-2 &  02 01 38.73 & --30 42 33.0  & 30/09/97  &  3$\times$1200 & 0.029 & 3$\times$400 & 0.036 & 2$\times$300 & 0.028 & 2$\times$300 & 0.033 &20.9 \\ 
f414-1 &  02 06 33.93 & --28 18 41.0  & 25/08/95  &  -- & -- & 1$\times$1200 & 0.027 & -- & -- & 1$\times$600 & 0.030 &21.2 \\ 
f415-1 &  02 20 47.83 & --31 54 15.1  & 24/08/95  &  -- & -- & 1$\times$1200 & 0.027 & -- & -- & 1$\times$600 & 0.040 &21.2 \\ 
f416-2 &  02 47 31.70 & --30 18 30.5  & 15/09/96  &  3$\times$1200 & 0.040 & 2$\times$600 & 0.042 & 2$\times$300 & 0.036 & 2$\times$300 & 0.038 &20.9 \\ 
f417-2$^*$ & 03 06 29.09 & --29 40 28.1  & 28/09/97  &  3$\times$1200 & 0.020 & 3$\times$400 & 0.022 & 2$\times$300 & 0.025 & 2$\times$300 & 0.028 &20.7 \\ 
f417-1 &  03 12 31.47 & --27 50 46.5  & 26/08/95  &  -- & -- & 1$\times$1200 & 0.043 & -- & -- & 1$\times$600 & 0.052 &21.2 \\ 
f466-2 &  21 54 17.07 & --29 50 48.0  & 15/09/96  &  3$\times$1200 & 0.041 & 2$\times$600 & 0.033 & 2$\times$300 & 0.035 & 2$\times$300 & 0.037 &20.8 \\ 
f466-1 &  21 56 02.40 & --29 19 29.9  & 24/08/95  & -- & -- & 1$\times$1200 & 0.084 & -- & -- & 1$\times$600 & 0.068 &21.1 \\ 
f467-1 &  22 09 07.17 & --27 24 38.5  & 26/08/95  & -- & -- & 1$\times$600  & 0.047 & -- & -- & 1$\times$300 & 0.083 & 20.7 \\ 
f467-2 &  22 13 13.45 & --30 03 07.2  & 27/09/97  &  3$\times$1200 & 0.024 & 2$\times$600 & 0.023 & 2$\times$300 & 0.038 & 2$\times$300 & 0.035 &20.9 \\ 
f468-1 &  22 32 26.17 & --31 12 12.1  & 25/08/95  &  -- & -- & 1$\times$600 & 0.027 & -- & -- & 1$\times$300 & 0.031 &20.9 \\ 
f468-2 &  22 44 16.60 & --30 02 13.8  & 28/09/97  &  3$\times$1200 & 0.036 & 3$\times$400 & 0.022 & 2$\times$300 & 0.034 & 2$\times$300 & 0.029 &20.8 \\ 
f469-1 &  23 05 43.34 & --30 39 01.0  & 25/08/95  &  -- & -- & 1$\times$1200 & 0.031 & -- & -- & 1$\times$600 & 0.032 &21.2 \\ 
f470-1 &  23 19 39.65 & --29 31 26.1  & 25/08/95  &  -- & -- & 1$\times$600 & 0.036 & -- & -- & 1$\times$300 & 0.048 &20.8 \\ 
f470-2$^*$ & 23 26 02.57 & --29 22 29.4  & 29/09/97  &  3$\times$1200 & 0.022 & 3$\times$400 & 0.020 & 2$\times$300 & 0.022 & 2$\times$300 & 0.028 &21.0 \\ 
f471-1 &  23 47 38.91 & --28 08 28.3  & 25/08/95  &  -- & -- & 1$\times$1200 & 0.028 & -- & -- & 1$\times$600 & 0.035 &21.2 \\ 
f471-2 &  23 48 58.85 & --30 01 19.0  & 30/09/97  &  3$\times$1200 & 0.030 & 3$\times$400 & 0.032 & 2$\times$300 & 0.033 & 2$\times$300 & 0.037 &21.0 \\ 
\hline
\label{phottab}
\end{tabular}
\end{center}
\end{table*}

Observations of the sequence fields were carried out using the
Tek\#3 CCD mounted at Cassegrain focus on the 0.9m telescope at Cerro
Tololo Inter-American Observatory (CTIO).  Data was obtained in three
separate observing
runs, from 23 -- 27  August 1995 (observers: A. Ratcliffe and
Q.A. Parker), 15 -- 20 September 1996 (observer: S.M. Croom) and 
25 September -- 2 October 1997 (observer: S.M. Croom).
The Tek\#3 CCD is a $2048\times2048$, thinned, AR coated device with
low read noise and good quantum efficiency in the $U$ band.  Readout
is controlled by 
the ARCON system which has four separate amplifiers, allowing quad
readout, which allows a quick readout time of $\sim40$s.  The pixel scale is
$0.396$ arcsec pixel$^{-1}$ giving a large $13.5'\times13.5'$ field of
view, allowing us to obtain magnitudes for a large number of stars 
in each frame ($\sim100$ with $B<21$).  The first observing run
produced sequences 
in $B$ and $R$ in 14 independent UKST survey areas.  These fields were
centred on areas with bright ($B<17$) galaxies in order to verify the
calibration of the Durham/UKST FLAIR Galaxy Redshift Survey
\cite{arat96}.  However, the
majority of area in each field could also be used to obtain stellar
sequences.  The second and third observing runs produced a total of 12
CCD sequences in $UBVR$.  For these runs the fields were chosen to
contain a broad and even range of stellar magnitudes.  All the fields
observed in the three runs are listed in Table \ref{phottab}.  The filters
used for these 
observations were the standard CTIO $UBVR$ $3\times3$ inch filters,
$BVR$ being similar to the Harris $BVR$ set, while the $U$ filter uses a
Copper
Sulphate solution blocker.  Magnitude zero-points, extinction and
colour terms relating the instrumental magnitudes to the standard
Johnson-Kron-Cousins system were obtained from frequent observations
of standard stars from the E-region standards at $\delta=-45^{\circ}$
\cite{mcbl89}.  These standards were also used to monitor the
quality of photometric conditions during each observing run.  All
data observed in the first run was taken in photometric conditions.
Only one night of the second run was photometric, so only data from
this night is included in this paper.  During the third observing run,
the beginning 
of the night of the 26/9 was photometric, as was most of the 28/9.
The night of the 30/9 was completely photometric, and shorter exposures
were taken on this night to validate the calibration in
fields f417-2 and f470-2.  $1\times600$s, $1\times300$s,
$1\times150$s and  $1\times150$s exposures were taken in these fields
in $UBVR$ respectively.  All observations of the sequence fields were
made at low airmass, ${\rm sec}(z)\leq1.34$, with most of the fields
being observed at ${\rm sec}(z)\leq1.1$.

\section{Data Reduction}

\begin{figure*}
\centering
\centerline{\epsfxsize=14.0truecm \figinsert{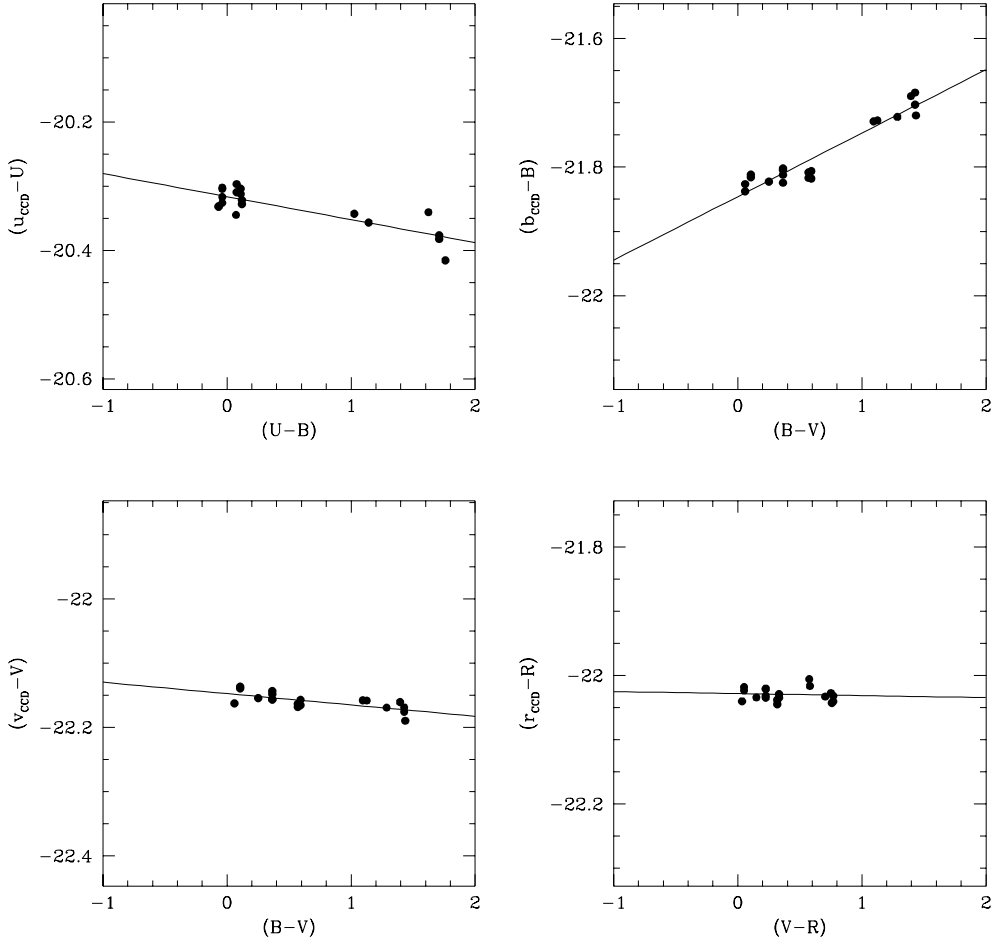}{0.0pt}}
\caption{Colour-magnitude relations in the $UBVR$ passbands derived
from observations of E-region photometric standards (Menzies et
al. 1989) during the 
1997 observing run.  We plot $m_{\rm CCD}-M$ (the extinction corrected
instrumental magnitude minus the known standard magnitude) against the
known colour of the stars.  The solid lines denote the best fit.}
\label{photfit}
\end{figure*}

Removal of the instrumental signature of the CCD from the images was
carried out using the {\small ARED.QUAD.QUADPROC} and {\small
IMRED.CCDRED.CCDPROC} tasks in the {\small IRAF} package.
{\small QUADPROC} is a variant of {\small CCDPROC} specifically designed to
reduce data 
from the multi-readout ARCON system.  This software removes the
instrumental bias and also corrects bad columns and pixels (of which
there are a number in the Tek\#3 CCD), as well as flat fielding each
frame.  In each passband we obtained both dome flats (taken during the
day) and sky flats (taken at twilight).  It was found that both
dome and sky flats adequately corrected small scale variations in
pixel sensitivity.  However, division by dome flats generally left a
residual large scale gradient across the image.  This gradient was at
the 5\% level in many cases.  Therefore, we used the sky flats,
which flattened the images at the 1\% level.  At least 5 separate
exposures were used to generate each flat-field.  These were median
combined using $3\sigma$ clipping to remove cosmic ray events (CREs).

We used the {\small IRAF} task {\small DIGIPHOT.DAOPHOT.PHOT} to
obtain photometry of all the E-field
standard stars observed.  We used 25-pixel ($10''$) radius apertures
centred on each standard star to obtain their instrumental magnitudes,
which we denote by $u$, $b$, $v$ and $r$;
the instrumental magnitude is simply given by
\begin{equation}
m=-2.5\log(f)+2.5\log(t),
\end{equation}
where $f$ is the flux and $t$ the exposure time.
The sky background value was obtained from the mode of an annulus
between 35 and 50 pixels in radius.  The {\small IRAF} task {\small
DIGIPHOT.PHOTCAL.FITPARAMS} -- an inter-active multi-parameter fitting
routine -- was used
to obtain the magnitude zero-points, colour and extinction terms.  For
the first set of data ($BR$ sequences) we assumed standard extinction
values for Cerro-Tololo: $A_{\rm B}=0.209$ mag airmass$^{-1}$, $A_{\rm
R}=0.108$ mag airmass$^{-1}$.  For these data the maximum airmass
reached was 1.341 in $r$ and 1.292 in $b$, with over 50\% of the
observations obtained at airmasses of less than 1.1.  Even in
the most extreme cases this assumption would only cause a zero-point error
of less than 0.01 mag.  For the second and third  sets of observations
($UBVR$ sequences) the extinction term was fitted as well as the
zero-point and colour terms.  We fitted inverse colour equations of
the form
\begin{equation}
b=B-X+Y(B-V)+A_{\rm B}{\rm sec}(z),
\end{equation}
where $X$ is the zero-point, $Y$ is the colour term, $A_{\lambda}$ is the
extinction value and $UBVR$ denote the standard magnitude
system.  The parameters from these fits are given
in Table \ref{photsol}, and example colour-magnitude relations from
the 1997 observing run are shown in Fig. \ref{photfit}.  The colour
terms are with respect to $(B-R)$ 
for both $B$ and $R$ in the 1995 observing run, while for the 1996 and
1997 runs they are  with respect to  $(U-B)$, $(B-V)$,  $(B-V)$ and
$(V-R)$ for $U$, $B$, $V$ and $R$ respectively.

\begin{table}
\baselineskip=20pt
\begin{center}
\caption{Photometric solutions (found using the {\small IRAF} routine
{\small FITPARAMS}) for standard stars observed in each of the three
observing runs.}
\vspace{0.5cm}
\begin{tabular}{cccccc}
\hline 
Obs.& Band & Extinction & Zero & Colour & Fit \\
Run & & mag airmass$^{-1}$ & Point & Term & r.m.s. \\
\hline 
1995 & $B$ & 0.209 & 22.442 & 0.095 & 0.003 \\
1995 & $R$ & 0.108 & 22.566 & 0.006 & 0.004 \\
1996 & $U$ & 0.453 & 20.240 & --0.035 & 0.014 \\
1996 & $B$ & 0.232 & 21.800 & 0.110 & 0.014 \\
1996 & $V$ & 0.131 & 22.030 & --0.019 & 0.007 \\
1996 & $R$ & 0.113 & 21.992 & --0.002 & 0.005 \\
1997 & $U$ & 0.459 & 20.316 & --0.036 & 0.017 \\
1997 & $B$ & 0.202 & 21.846 & 0.099 & 0.016 \\
1997 & $V$ & 0.138 & 22.147 & --0.018 & 0.009 \\
1997 & $R$ & 0.066 & 22.028 & --0.003 & 0.010 \\
\hline 
\label{photsol}
\end{tabular}
\end{center}
\end{table}

\begin{figure}
\centering
\centerline{\epsfxsize=8.5truecm \figinsert{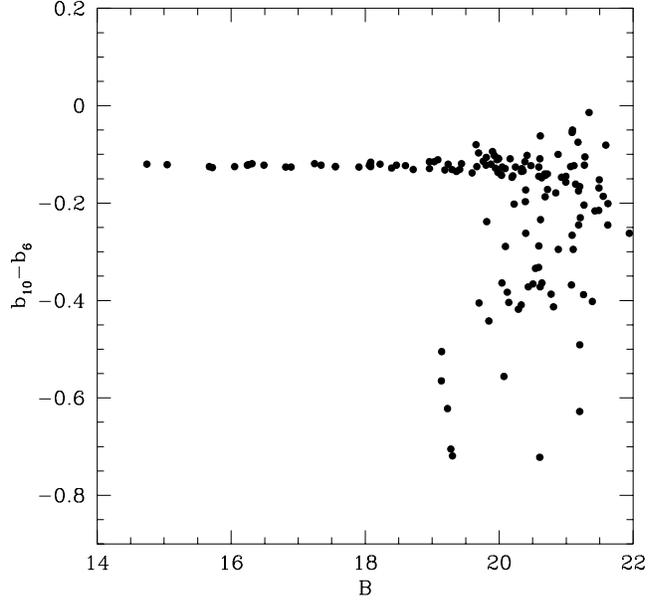}{0.0pt}}
\caption{The star-galaxy separation parameter, $b_{10}-b_{6}$, plotted
as a function of $B$ for field f471-2.  Objects with large negative
values are galaxies, with the divide between stellar and non-stellar
sources being at $b_{10}-b_{6}\sim-0.2$.}
\label{sg}
\end{figure}

The sequence fields were combined using {\small IRAF}'s {\small
IMCOMBINE} task, with the {\small
CRREJECT} algorithm used to reject CREs.  In the case of single
exposures on a field (as was the case for data from the first run), we
used the image detection routines, and the matching of images in
different bands to  reject CREs.  All  images with peak
counts higher than a $5\sigma$ threshold above the sky background were
selected using {\small DAOFIND}.  Aperture photometry was obtained for
these objects using {\small PHOT} within 10 and 25 pixel apertures.
The larger aperture, corresponding to $10''$ radius, is the same as that used
for our standard observations and defines the zero-point for our
magnitude system.  The median seeing during the observations was
$1.5''$ and was rarely above $2.0''$.  At faint fluxes the sky counts add
considerable noise within an aperture of this size.  For this reason
we use the magnitude obtained within a 10 pixel aperture and then
corrected it
to the 25 pixel magnitude scale using the brightest
unsaturated stars (at least 8 per field) with no close companions in
each field to define the aperture correction.  This corrections was typically
$\sim0.03$ mag  (see Table \ref{phottab}), with an r.m.s. error of
$\sim0.002$ mag.  We also obtained aperture photometry within a 6 pixel
aperture.  The difference between this and the 10 pixel aperture
magnitude was used to give an approximate star--galaxy separation
criterion.  Fig. \ref{sg} shows this plotted as a function of $B$ magnitude. 

\begin{figure*}
\centering
\centerline{\epsfxsize=14.0truecm \figinsert{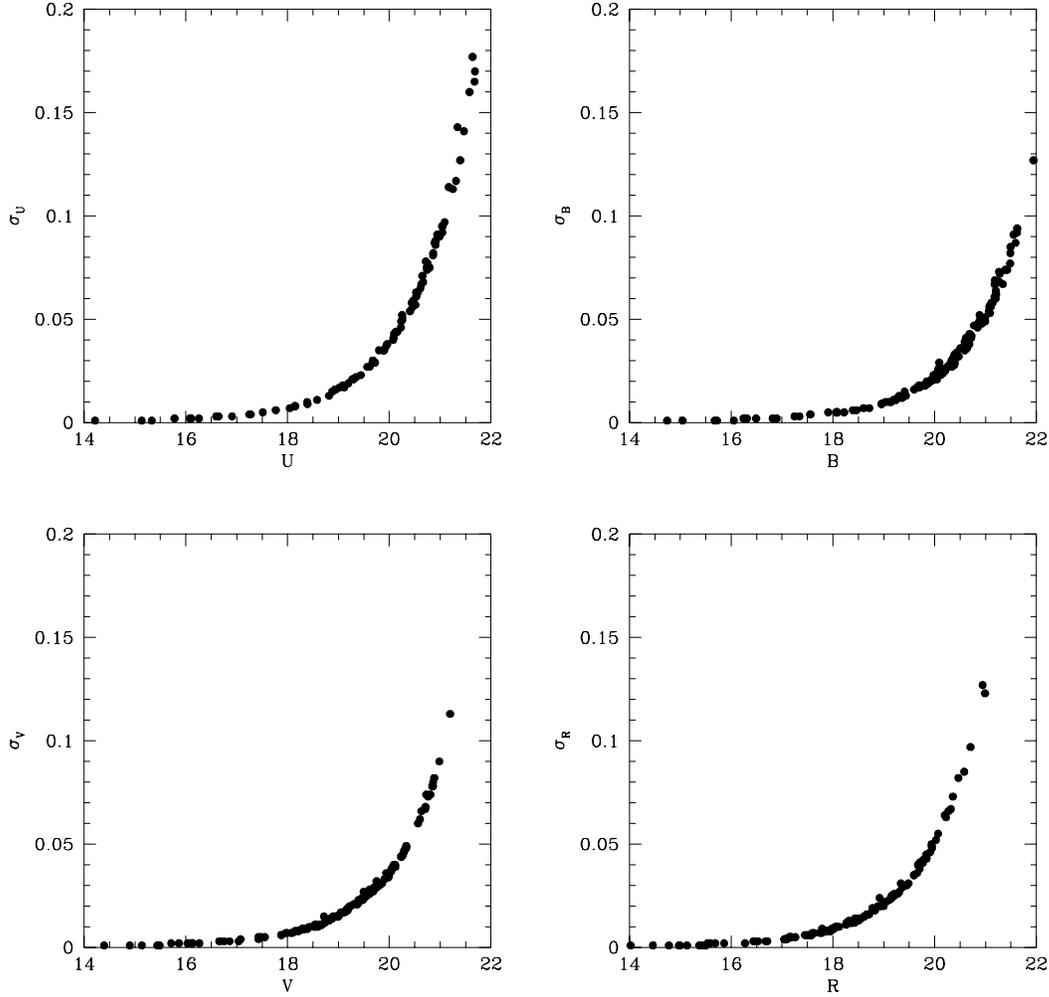}{0.0pt}}
\caption{Photometric errors as a function of magnitude for the fully
reduced and calibrated data in field f471-2.}
\label{photerrs}
\end{figure*}

\begin{figure*}
\centering
\centerline{\epsfxsize=15.0truecm \figinsert{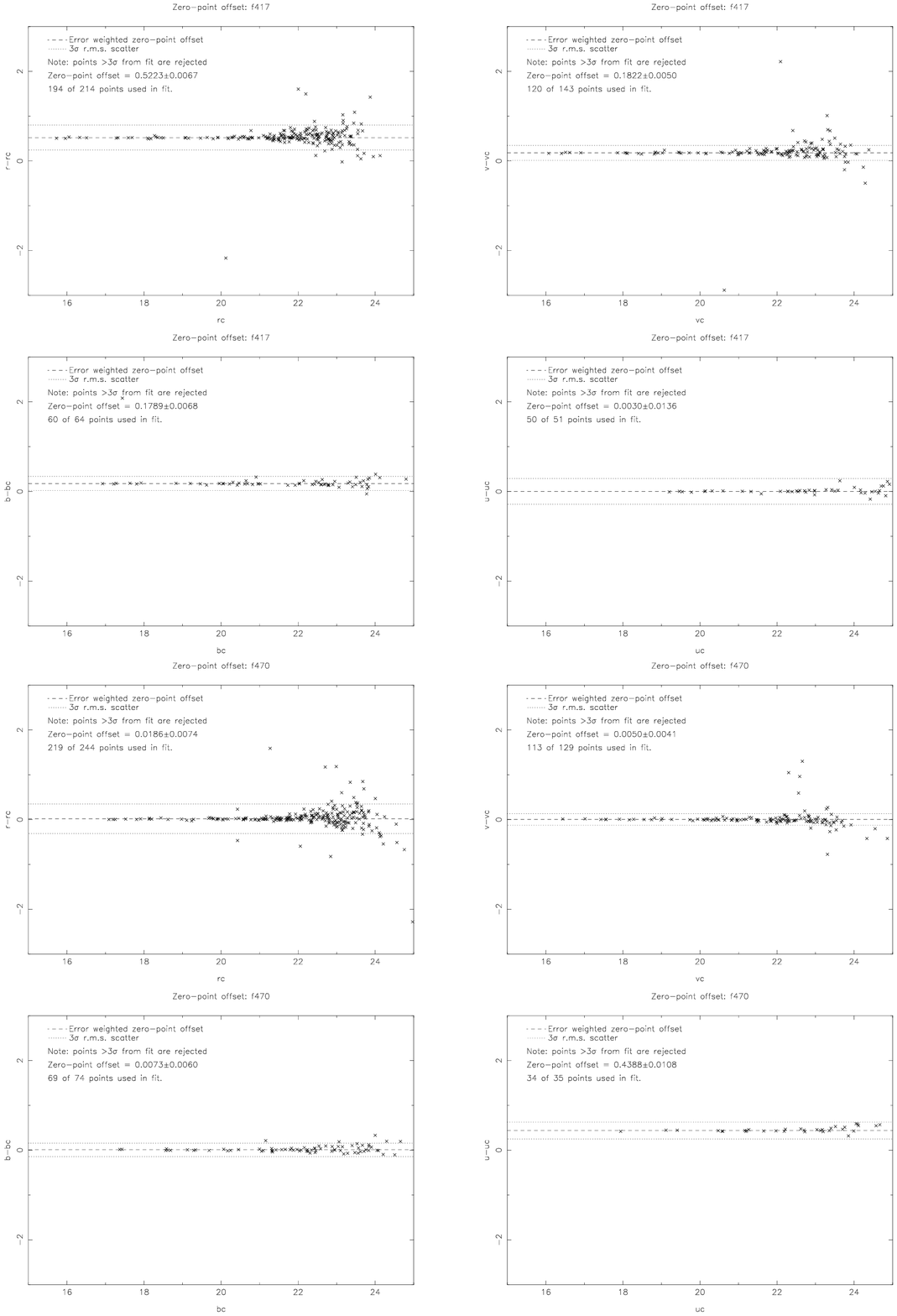}{0.0pt}}
\caption{Zero-point corrections for data observed in non-photometric
conditions (f417-2 and f470-2).  We plot the difference between the
photometric (e.g. $rc$) and non-photometric (e.g. $r$) instrumental
magnitudes in each band.  The measured offset is shown, along with the
$3\sigma$ upper and lower bounds.}
\label{zpcalib}
\end{figure*}

The aperture corrected
instrumental magnitudes were then converted to the standard $UBVR$ system
using the {\small INVERTFIT} task and the colour equations derived above.
The Poisson photometric errors were plotted as a function of magnitude
(an example of which is shown in Fig. \ref{photerrs}) in order to
estimate the effective limiting 
magnitude of the standard observations in each field.  We define this
limit to be the 
magnitude at which the photometric errors reach 0.05 mag.  These
limiting $B$ band magnitudes are shown in Table \ref{phottab}.

Two fields, f417-2 and f470-2, were observed in non-photometric
conditions.  We obtained shorter exposures in these fields in an
attempt to measure an accurate zero-point.  Fig. \ref{zpcalib} shows
the offsets between the photometric data (e.g. $rc$) and
non-photometric data (e.g. $r$).  We fit a constant to any observed offset,
rejecting points that are $>3\sigma$ from the fit in an iterative
process.  We find that there is no significant offset in the $u$ data
in f417-2, while the $bvr$ data does show a significant offset.  For
f470-2, the $u$ photometry shows the only large offset, of $\simeq0.4$
mag.  The offsets and errors on the offsets are listed in Table
\ref{offsets}.  While these data can in principle be used in a similar
manner to the other sequences, any potential users should note that the
error in the zero-point offset will increase significantly the total
photometric errors.  For example, the $U$ photometry in field f417-2
would then have a minimum error of $0.022$ mag.

\begin{table}
\begin{center}
\caption{Zero-point offsets for non-photometric fields.}
\vspace{0.5cm}
\begin{tabular}{cccc}
\hline 
Field & Band & Offset & $\sigma$ \\
\hline 
f417-2 & $r$ & 0.5223 & 0.0067\\
f417-2 & $v$ & 0.1822 & 0.0050\\
f417-2 & $b$ & 0.1789 & 0.0068\\
f417-2 & $u$ & 0.0030 & 0.0136\\
f470-2 & $r$ & 0.0186 & 0.0074\\
f470-2 & $v$ & 0.0050 & 0.0041\\
f470-2 & $b$ & 0.0073 & 0.0060\\
f470-2 & $u$ & 0.4388 & 0.0108\\
\hline 
\label{offsets}
\end{tabular}
\end{center}
\end{table}

The brightest 10 unsaturated stars in each field were matched to
their positions on APM scans of UKST III-aJ survey plates which
in turn have been matched to the PPM (B1950) \cite{ppm93} astrometric
catalogue.  These stars were then used to derive a transform from CCD
$(x,y)$ positions to $(\alpha,\delta)$ using the Starlink {\small
ASTROM} software.

\section{Results}

\begin{figure*}
\centering
\centerline{\epsfxsize=14.0truecm \figinsert{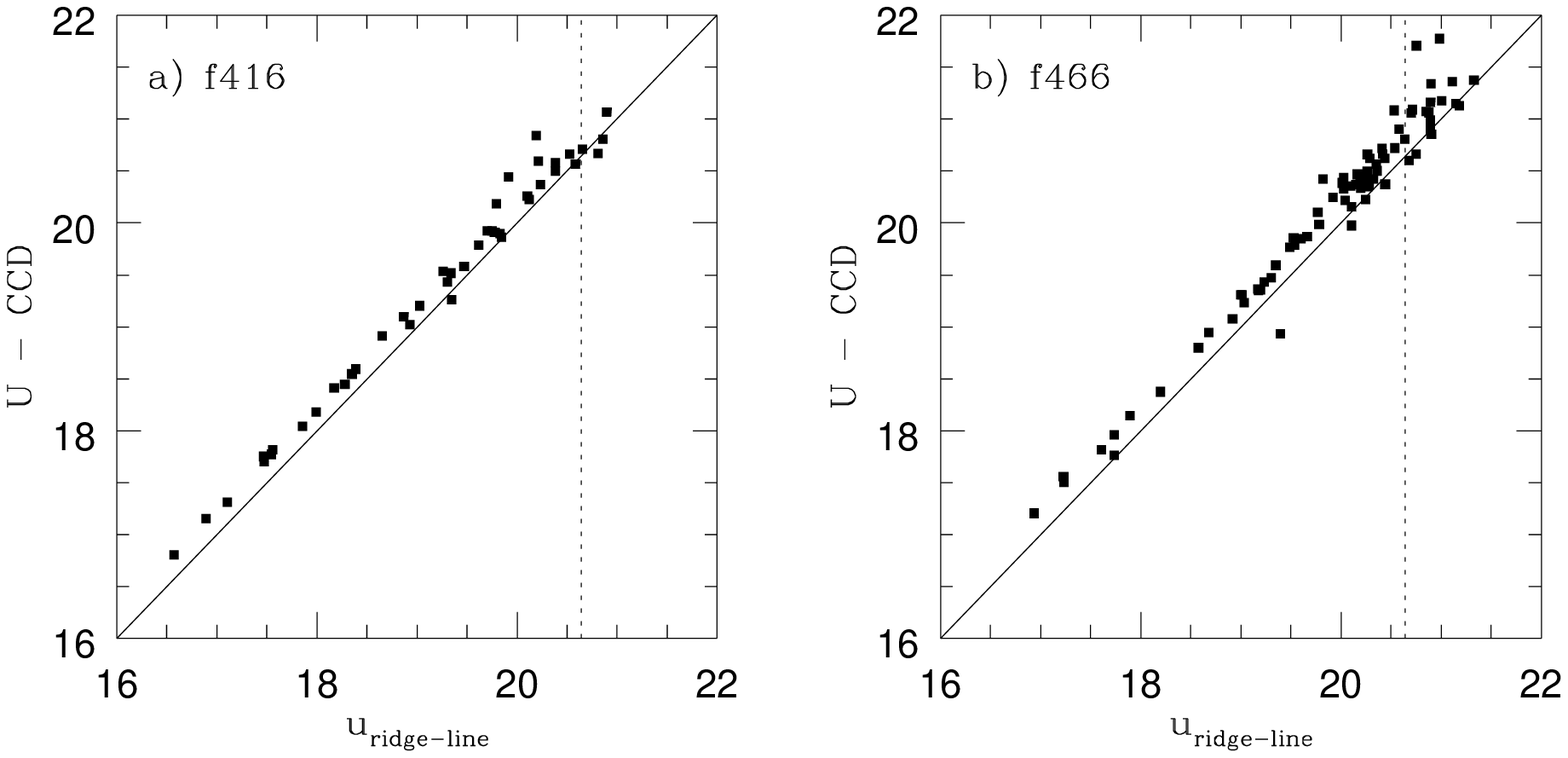}{0.0pt}}
\caption{An example of the $U$ photometry from this paper used to
check the calibration in the 2dF QSO Redshift Survey for fields a)
f416 and b) f466.  We plot $U$ CCD
photometry against photographic $u$ photometry from UKST plates
calibrated by boot-strapping to the $u-b_{\rm J}$ colours of main
sequence stars.  The dotted vertical line shows the effective
magnitude limit of the survey in the $U$ band.}  
\label{qsouphot}
\end{figure*}

In Tables 4--29 (available electronically) we list the positions,
$UBVR$ magnitudes and
Poisson errors for all our sequence stars.  We note that the errors
listed in these tables are purely Poisson errors and do not account
for systematic uncertainties.  Objects which have Poisson photometric
errors smaller than 0.01 mag have not been given an error, as
it is clear that at bright magnitudes the
Poisson error estimate breaks down and the errors are dominated by
systematic effects such as the accuracy of the determination of
zero-points, colour terms and aperture corrections.  The
photometric fits are clearly the main source of error, as the typical
aperture correction errors were $\sim0.001$ mags and were rarely above
$\sim0.003$ mags.  Therefore, the r.m.s. errors from the fits listed in
Table \ref{photsol} should be taken as lower limits to the total
photometric errors.  Also listed are the 25
pixel aperture magnitudes, which give good estimates of total galaxy
magnitudes for galaxies with isophotal diameters less than $20''$.
The star galaxy separation parameter, $b_{10}-b_{6}$, is given and
generally objects with $b_{10}-b_{6}<-0.2$ will be galaxies, although
this does vary somewhat from field to field, depending on the seeing
in each observation.  The Poisson errors on the aperture corrected
10 pixel aperture magnitudes is $\sim0.05$ mag at $B=21$ (see
Table \ref{phottab}), while the 0.05 mag error limit for the other
bands is typically 20.2, 20.5, 20.0 for $U$, $V$ and $R$ respectively
(Fig. \ref{photerrs}).  We only list objects for which the
photometric errors in $B$ are $\leq0.05$ mag.

The photometric sequences presented in this paper are intended
primarily for the calibration of photographic plates.  In this case,
it is expected that the sequences be used as a whole to obtain a
calibration curve, rather than to provide accurate magnitudes
for individual stars.  As we have not repeated our observations over a
sufficient time scale we are unable to remove variable objects from
our catalogue.  Also, as discussed in Paper I, the most extreme
sequence stars have colours outside of the range of the standards used
(namely, $-0.068<U-B<1.760$, $0.060<B-V<1.436$ and $0.035<V-R<0.770$).
The colour terms in Table \ref{photsol} are small (the $B$ colour term
is by far the largest at $\sim0.1$), therefore some extrapolation
beyond the colour range of the standards should not have a serious
effect on photometry.  However, we note that due to potential
non-linearities of the colour terms (particularly in $R$ for extremely
red colours) outside of the range of the standards, potential uses
should be wary of extreme extrapolation \cite{bessell90}.
Of course, it is clearly possible to construct sequences from our
data containing only stars which do have colours within the range of the
standard observations.  The three observing runs also give us
independent measurements of the colour terms which should not have
changed from run to run.  We have three measurements of the colour
terms for $B$ and $R$, with the scatter between them being 0.008 and
0.005 respectively.  We have only two measurements of the $U$ and $V$
colour terms, but in both cases the difference between the 1996 and
1997 measurements was only 0.001.  This consistency between runs gives
us great confidence in the reliability of our measurements.

These sequences and the others from this project have already been used in
several wide area surveys, including the 2dF QSO Redshift Survey
\cite{smith96} and the 2dF Galaxy Redshift Survey \cite{maddox98}
amongst others.  In Fig. \ref{qsouphot} we demonstrate possible uses of
this data set by way of a photometric calibration example taken from 
Croom 1997.  The 2dF QSO Redshift Survey candidate catalogue is 
constructed from $u$, $b_{\rm J}$ and $r$ UKST plates.  The
photographic $u$ photometry was calibrated by boot-strapping from
available $B$ CCD photometry via the colour of the main-sequence stellar
ridge-line.  In practice, this is done by straightening the stellar locus in
the $u-b_{\rm J}$ vs. $u$ plane.  In Fig. \ref{qsouphot} we compare
the ``ridge-line'' photometry to data from our CCD sequences f416-2
and f466-2.  This figure shows that the boot-strapping has indeed linearized
the photographic magnitudes.  However, the process leaves a $\sim0.2$
mag offset with respect to the CCD photometry.  Of course, this has no
practical effect on the QSO selection as the candidate selection
criteria are determined relative to the stellar locus.  

We hope that
by making this data available to the general community it will help in
the photometric calibration of many other wide-field projects.

\section*{Acknowledgements}

SMC acknowledges the support of a Durham University Research
Studentship.  AR and RJS acknowledge the support of PPARC Studentships.
The data was reduced using the Durham STARLINK node.  We thank CTIO
for supporting this project and thank the staff at CTIO for their
invaluable help with our observations.
\nocite{smith98}

{}

\end{document}